\documentclass[a4paper,12pt]{article}

\usepackage{amsmath}\usepackage{amssymb}\usepackage[dvips]{graphicx}
\usepackage{latexsym}\usepackage{cite}\usepackage{bbold}

\usepackage{pstricks}
\usepackage{dcolumn}
\usepackage{bm}

\newcommand{\be}{\begin{equation}}
\newcommand{\ee}{\end{equation}}

\newcommand{\ka}{\kappa}

\newcommand{\mcF}{{\mathcal F}}

\newcommand{\mcK}{{\mathcal K}}
\newcommand{\mcL}{{\mathcal L}}
\newcommand{\mcN}{{\mathcal N}}

\newcommand{\mcW}{{\mathcal W}}
\newcommand{\om}{{\omega}}

\def\beq{\begin{equation}}
\def\eeq{\end{equation}}
\def\pl{\partial}

\def\al{\alpha}
\def\bt{\beta}
\def\Ga{\Gamma}
\def\ga{\gamma}

\def\De{\Delta}

\def\ka{\kappa}
\def\si{\sigma}
\def\Si{\Sigma}
\def\te{\theta}
\def\La{\Lambda}
\def\lam{\lambda}
\def\Om{\Omega}
\def\om{\omega}
\def\ep{\epsilon}

\def\sq{\sqrt}

\def\l{\left (}
\def\r{\right )}

\def\fr{\frac}
\def\la{\label}
\def\hs{\hspace}
\def\vs{\vspace}

\def\ran{\rangle}
\def\lan{\langle}
\def\ov{\overline}
\def\tl{\tilde}

\begin{document}
\begin{titlepage}
\begin{flushright}
CERN-PH-TH-2006-141\\
HD-THEP-06-14\\
August 8, 2006
\end{flushright}
\vspace{0.6cm}
\begin{center}
{\Large \bf Radion Stabilization In 5D SUGRA}
\end{center}
\vspace{0.5cm}

\begin{center}
{\large
Filipe Paccetti Correia$^a$\footnote{E-mail address:
paccetti@fc.up.pt},
Michael G. Schmidt$^b$\footnote{E-mail address:
m.g.schmidt@thphys.uni-heidelberg.de},
Zurab Tavartkiladze$^c$\footnote{E-mail address:
zurab.tavartkiladze@cern.ch}}

\vspace{0.3cm}

$^a${\em
Centro de F\' isica do Porto,
Faculdade de Ci\^ encias da Universidade do Porto\\
Rua do Campo Alegre 687, 4169-007 Porto, Portugal\\

$^b$
Institut f\"ur Theoretische Physik,
Universit\"at Heidelberg\\
Philosophenweg 16, 69120 Heidelberg, Germany\\

$^c$ Physics Department, Theory Division, CERN, CH-1211 Geneva 23, Switzerland
}
\end{center}
\vspace{0.4cm}
\begin{abstract}

We present a detailed study of radion stabilization
within 5D conformal SUGRA compactified on an $S^{(1)}/Z_2$ orbifold.
We use an effective 4D superfield description developed in our previous work.
The effects of  tree level bulk and boundary couplings, and in particular
of one loop contributions and of  a non perturbative correction on the radion
stabilization are investigated.
We find new examples of radion stabilization in non SUSY
and (meta-stable) SUSY preserving Minkowski vacua.

\end{abstract}

%
%

\end{titlepage}

\vs{0.3cm}

\section{Introduction}

The stabilization of moduli is a central problem if one is
searching for vacua of string theory and their possible realization in
the real world. Models with one extra dimension allow to study this and other
questions in a simplified setting. They might lead to interesting
insights in physics \cite{Randall:1999ee}, e.g. concerning the generation of
hierarchies, the discussion of SUSY breaking, or a new look to
inflation in early cosmology. They even may be directly related to
string/M-theory in an intriguing way, if Calabi-Yau and 5D
compactification differ in scale. Last not least 5D supergravity is an
(admittedly still technically complicated but) in principle well
understood field.

{}For supergravity with a fifth dimension on a $S^{(1)}/Z_2$ orbifold
off-shell SUGRA (i.e. with auxiliary field components, not integrated out)
certainly is a very good choice. This was pioneered in \cite{Zucker:1999ej}.
In our line of research \cite{us04a,Correia:2004pz, Correia:2006pj } we rely on the work of
Fujita, Kugo, Ohashi \cite{Fujita:2001bd} on
5D superconformal gravity including also vector and hypermultiplets and
using compensators of that type. We have reformulated important parts of
this work\footnote{What is still lacking is a superfield formulation of
the odd part of the Weyl multiplet and of related covariant derivatives.
This would be
needed for a genuine loop calculation containing gravitational higher KK
modes in super language like in \cite{Buchbinder:2003qu}.} in terms of a
4D superfield formalism (see also \cite{Abe:2004ar} and subsequent
developments \cite{AbeS2}) including
vector and chiral multiplets and 4D supergravity. In a recent paper
we \cite{Correia:2006pj} succeeded to integrate out a multiplier real
superfield ${\mathbb W}_y$
contained in the even part of the 5D Weyl multiplet. Thus we obtained a
superspace action where the radion field $e^5_y$ is now contained only in
chiral superfields. Depending on the problem one wants to attack one can
gauge fix the superconformal symmetries already in the 5D theory or only
in the effective 4D theory which turns out to be still 4D superconformal
invariant in this version.

There have been numerous  studies of moduli (radion) stabilization
(see for instance \cite{luty99}-\cite{Villadoro:2005yq}, \cite{Correia:2006pj}
and references therein) and various mechanisms were suggested.
In this work we consider this issue within our formalism which greatly
simplifies investigations. Also, new examples of moduli stabilization are
presented.

The paper is organized as follows:
Starting point in sect. \ref{sec:reduction} is a brief
presentation of our formalism developed further in our recent work
\cite{Correia:2006pj}. In sect. \ref{sec:stabil} the issue of moduli
stabilization is discussed
and examples of stabilization due to 1-loop corrections and non perturbative
effects are presented. In section \ref{sec:treeuplift} we discuss
the uplift and
stabilization in SUSY breaking and SUSY preserving (meta-stable)
Minkowski ($M_4$)
vacua. In sect. \ref{sec:spectrum} we present some mass formulas and
estimates. We also mention some possible phenomenological implications
of considered scenarios.
In sect. \ref{sec:conc} we conclude with a short summary.
Appendix \ref{appA} deals with calculational tools of the various 1-loop
corrections to the K\"ahler potential.

\section{Effective 4D Superconformal Description}\label{sec:reduction}

We use an effective 4D superconformal description
of 5D SUGRA in which case the gauge
and hypermultiplet action can be built by knowing the forms of the K\"ahler
potential $\mcK$, the superpotential $W$ and the gauge kinetic function
$f_{IJ}$. The Lagrangian couplings are

\be
          \mcL_{D}^{(4D)}=- 3\int d^4\theta \,e^{-\mcK/3}\,\phi^+\phi ~,
\la{eleven}
\ee

\be\label{eq:sup}
\mcL_W^{(4D)}=\int d^2\theta \phi^3 W~,
\ee

\be
\mcL_V^{(4D)}=\frac{1}{4}\int d^2\theta
\,f_{IJ}(\Sigma){\tilde\mcW}^{\alpha I}
{\tilde\mcW}_{\alpha}^J+\textup{h.c.}
\la{gaugekin}
\ee
where $\phi $ is the 4D conformal compensator, a chiral superfield.

If the theory under discussion is 5D orbifold SUGRA, then as derived in
\cite{Correia:2006pj}, at tree
level we have the K\"ahler potential
\beq
{\cal K}=-\ln \tl{\cal N}(\Si +\Si^{\dagger })-
2\ln \l 1-H^{+}e^{-g_IV^I}H\r ~,
\la{genK}
\eeq
where $\Si $ is a modulus emerging from a 5D gauge supermultiplet and
$H$ is a chiral superfield which is part of the 5D hypermultiplet and
which is charged under the 4D vector superfields
$V^I$ (here we assume that the hyperscalar manifold is the simplest
one). The ${\tilde\mcW}$ denotes the 4D gauge field-strength superfield, while
$f_{IJ}(\Sigma)=-{\hat\mcN}_{IJ}(\Sigma)$
with double derivative  $\hat\mcN_{IJ}$ obtained
from the $3^{\rm rd}$ order norm function
$\mcN(\Sigma)$ by setting at the end the odd fields to zero
(more relations between 4D and 5D objects are given below).

The scalar potential derived from (\ref{eleven})-(\ref{gaugekin})
consists of two parts: the F-term potential is given in terms of
the K\"ahler potential and the superpotential as
\be\label{eq:formula}
                 V_F=M_{\rm Pl}^4\,e^{\mcK}\left(\mcK^{I{\bar J}}D_IWD_{\bar J}{\bar W} - 3|W|^2\right), \quad \textup{with } D_I\equiv\partial_I+\mcK_I
\ee
where $I$ runs over chiral multiplets emerging from both hyper and
vector scalars ($I=H,\Sigma$).
It is easy to write down also the $D$-term potential
\beq
V_D=-\fr{1}{4}M_{\rm Pl}^4\hat{\cal N}_{IJ}{\rm D}^I{\rm D}^J~,~~~~
{\rm with}~~~
{\rm D}^I=-\hat{\cal N}^{IJ} g_{4J}q_i{\cal K}_{H_i}H_i ~,
\la{Dpot}
\eeq
where we took into account that the ${\cal K}$ includes the hypermultiplets
as ${\cal K}(H_i^{\dagger }e^{-q_ig_{4I}V^I}H_i)$ (here we have restored the
$U(1)$ charge $q_i$ of the chiral superfield $H_i$).
Eq. (\ref{Dpot}) does not contain contributions from
FI terms arising if the compensator hypermultiplet is charged under
$U(1)_R$.
$D$-terms may play crucial role for uplifting to the Minkowski
or de-Sitter vacua \cite{Villadoro:2005yq}.
In this work we will not consider  $D$-term potentials but
instead will investigate the role of $F$-term potentials (of eq.
(\ref{eq:formula})) in the radion stabilization.

Our investigation will focus mostly on the \emph{vector} moduli
$\Sigma$, as the \emph{hyper} moduli $H$ can be stabilized in a fairly
simple way by using tree-level brane potentials.
Stabilization of vector moduli is more complicated because they do not
couple directly to the branes.
Here indeed lies the difficulty of radion stabilization. Recall that
the radion  (the $e^5_y$ component of 5D Weyl multiplet) is related to
the chiral superfields $\Si^I_5$
\be
\Si^I_5=\fr{1}{2}(e^5_yM^I+iA_y^I)+\cdots
\eeq
Since for this paper we consider unwarped geometries, the zero modes of the
various superfields are $y$-independent. In the
following we consider a dimensionless $e^5_y$ and a
fixed length $R$  ($-\pi R<y<\pi R$) is introduced. Note
that $R$ is not dynamical, but for practical reasons we will take it to
coincide with the size of the extra-dimension (after the latter is
stabilized).

Before closing this section we present relations between $5D$ and $4D$ objects
and give some explanation for the notations used in the paper.
The fields $\Si $, $H$ in eq. (\ref{genK}) and appearing in 4D Lagrangian
densities are dimensionless and are related to the corresponding 5D
components as
\beq
\Si =\ka_{5}\Si_5~,~~~H=\ka_{5}H_5~,~~~~{\rm with}~~~\ka_5=M_5^{-3/2}~,~~
M_{\rm Pl}^2=2\pi RM_5^3~.
\la{4Drel5D}
\eeq
Moreover, we have
\beq
\tl{\cal N}=\ka_{5}^2{\cal N}~,~~~{\tilde\mcW}=\sq{2\pi R}{\mcW}~,~~~
\sq{2\pi R}g_4=g_5~.
\la{4Drel5Dmore}
\eeq
At some places (for instance in eqs. (\ref{eq:hyper_phys_charged_1}),
(\ref{sup_odd_g}), (\ref{evCoup}), (\ref{oddCoup}))
we also present 5D Lagrangian couplings. In order to make expressions more
compact  there we omit subscript $5$  for the 5D objects, hoping that this
will not create any confusion.

\section{Moduli Stabilization}\label{sec:stabil}

We start our discussion with a K\"ahler potential including gauge moduli only.
At tree level, due  to the no-scale nature of the $\Sigma$,
which means that at tree-level
$\mcK^{\Sigma{\bar\Sigma}}\mcK_{\Sigma}\mcK_{\bar\Sigma}=3$, and with moduli
independent superpotential $W={\rm constant}$, from (\ref{eq:formula}) we
see that a vector modulus is a flat direction and $V_F=0$. Inclusion of
moduli dependent tree level superpotentials (consistent with gauge symmetries)
can lead to moduli stabilization however in an AdS vacuum
in simple minded cases \cite{Correia:2006pj}.
The vacuum uplifting is an important issue and different mechanisms can be
applied. Later in this  paper we will use and discuss some of them.

As far as the hypermultiplet $H$ moduli ('hypers') are concerned,
they can be stabilized in a simpler way. The brane localized superpotential
$W(H)$ can fix the value of $H$. Apart from this, hyper moduli generate
positive definite contribution to the potential

\be
V_F(H)=M_{\rm Pl}^4\,e^{\mcK}\mcK^{H{\bar H}}D_HWD_{\bar H}{\bar W},
\ee
which can also serve for a self consistent uplift as we will discuss in
more detail in section \ref{sec:treeuplift}.

\subsection{Stabilization and uplift by 1-loop corrections}
\label{subsec:kaehlerstuff}

As originally proposed in \cite{Ponton:2001hq,Luty:2002hj}, one-loop
corrections to the K\" ahler potential, due to bulk multiplets, can lead
to moduli stabilization in $M_4$ vacuum. A more detailed discussion of
this kind of stabilization mechanism was then presented in
refs.\cite{vonGersdorff:2003rq,dudas05} (see \cite{vonGersdorff:2005ce} for a
discussion of two-loop effects), for the one modulus (i.e. the radion) case.
Here we will present the supergravity embedding of these models, generalizing
them to the many moduli case, and consider \emph{new} one-loop corrections
to $\mcK(\Sigma,\Sigma^+)$. The new corrections will be crucial to remove
the flatness of the potential in the axionic directions, as they break the
continuous \emph{shift symmetry} present at tree-level.

There are loop corrections to the K\"ahler potential which may be either
moduli-independent or moduli-dependent. Moduli dependence can appear in a
combination which measures the size of the extra dimension, the radion.
Moduli also may appear in different combinations. The latter happens when
some multiplets are charged under some bulk abelian gauge symmetries
(that is, some isometries of the scalar manifolds are gauged by vector
multiplets). These two kinds of corrections correspond to the contributions
of massless and massive bulk multiplets of rigid supersymmetry of
refs.\cite{Luty:2002hj,vonGersdorff:2003rq}.

In our studies we will include moduli-independent corrections coming in
general from bulk Weyl, vector and hyper supermultiplets. For the generation
of moduli dependent corrections we will introduce bulk hypermultiplets. As
we will see shortly, each type of correction will be important for moduli
stabilization in Minkowski vacuum. Let's consider these 1-loop corrections
to the K\"ahler potential in more detail.

\begin{center}
{\bf Moduli-independent one-loop contributions}
\end{center}

We consider first the contributions of the gravitational sector, of abelian
vector multiplets in the unbroken phase, and of uncharged hypermultiplets.
These lead to a 1-loop correction to the low-energy D-term Lagrangian of the
form,
\be
\Delta \mcL_V=-\int d^4\theta\frac{\alpha}{R^3{\mathbb W}_y^2}+
(\textup{higher powers in a super-der. expansion}),
\la{invW}
\ee
where $\alpha=(-2-N_V+N_H)\frac{\zeta(3)}{2(2\pi)^5}$ depends on the number
$N_V$ of "massless" vector multiplets and on the number $N_H$ of "massless"
hypermultiplets. Note that (\ref{invW}) is a 5D Lagrangian coupling.
Integrating out ${\mathbb W}_y$ \cite{Correia:2006pj} one finds
(still in the 5D theory) an effective 4D one-loop K\" ahler potential
\be\label{eq:kaehler_with_corrections}
\mcK(\Sigma,\Sigma^+)= -\ln\left({\tilde\mcN}(\Sigma+\Sigma^+)+\Delta\right),
\ee
where
\be\label{eq:kaehler_corrections_massless}
                   \Delta =\De_{\al }\equiv\ka_5^2\frac{\alpha}{R^3}.
\ee
Note that here the $N_H$ dependent part, arising by integrating out the
hypermultiplets, can be obtained from the cases presented below by taking
the superpotential  coupling of moduli with the hypermultiplets to be
zero (setting $g_{\bt }=0$ in eq. (\ref{eq:Delta_hyper_charged_1})).

\begin{center}
{\bf Moduli-dependent one-loop contributions}
\end{center}
There are several ways of gauging isometries of the scalar manifold and
therefore there are also different possible corrections to
$\mcK(\Sigma,\Sigma^+)$:

Let us start with a physical 5D hypermultiplet ${\mathbb H}=(H,H^c)\sim(+,-)$
charged under the vector multiplet ${\mathbb V}=(V,\Sigma)\sim(-,+)$, where
the parities under $S^{(1)}/Z_2$ orbifolding are shown in brackets.
The relevant coupling is \cite{us04a}
\beq
\int d^2\theta \,(2H^c\partial_y H-g_{\bt }\Sigma (H^2-H^{c2}))+
{\rm h.c.}~.
\label{eq:hyper_phys_charged_1}
\eeq
Note again that eq. (\ref{eq:hyper_phys_charged_1}) is a 5D Lagrangian density
(it should not be confused with the effective 4D description of
(\ref{eleven})-(\ref{gaugekin})) and all fields and coupling is five
dimensional.
This form is useful as a starting point when making loop calculations.
The corresponding one-loop correction due to hypermultiplets running in the
loop, obtained in appendix \ref{appA1}, is
\beq
\label{eq:Delta_hyper_charged_1}
\Delta_{\bt }=\ka_5^2\frac{\beta}{R^3}
\sum_{k=1}^{\infty }\fr{1}{k^3}
(e^{-k2\pi Rg_{\bt }\Sigma}+e^{-k2\pi Rg_{\bt }\Sigma^{+}})
\left(1+k\pi Rg_{\bt }(\Sigma+\Sigma^{+})\right)~,
\eeq
with $\beta=N_{H_{\bt }}/{4(2\pi)^5}$, where $N_{H_{\bt }}$ is the number of
charged hypermultiplets.
There are several ways of deriving eq.\eqref{eq:Delta_hyper_charged_1},
the most elegant one using the relation of the K\"ahler potential of
$\mcN=2$ supergravity to the prepotential. Another way is to make a KK
reduction and then to use the standard tools in order to evaluate the 1-loop
correction to the K\"ahler potential.
These two ways are presented in appendix \ref{appA1}.
For our purposes the higher powers in the superderivative expansion do not
play an important r\^ ole, we will therefore discard them in the following.
Another feature which we did not display is that
\eqref{eq:Delta_hyper_charged_1} is only valid for $Re(\Sigma)>0$, for
$Re(\Sigma)<0$ we must replace $\Sigma\to-\Sigma$. This expression leads to
new contributions to $\mcK(\Sigma,\Sigma^+)$ in
eq.\eqref{eq:kaehler_with_corrections}. In the most general case
$g_{\bt }\Sigma$ can be replaced by a linear combination of moduli
$g_I\Sigma^I$. \\

%
%
\begin{figure}[t]
\begin{center}
\resizebox{0.6\textwidth}{!}{
  \includegraphics{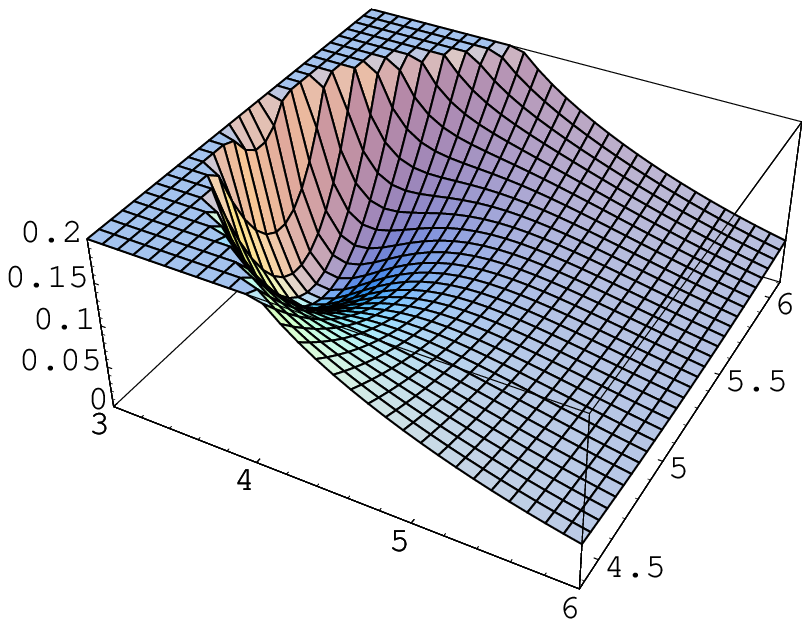}
}
\put(-205,35){$Re(\Si )$}
\put(-33,70){$Im(\Si )$}
\put(-320,100){$V/V_0$}
\caption{Effective potential as a function of $Re(\Si )$ and $Im(\Si )$,
for $\bar \al =0.437$, $\bar \bt =0.5$, $\bar \ga =1.5$ and $b=0.3$, $c=1$.
$V_0=8.5\cdot 10^{-6}M_{\rm Pl}^4|W|^2$.}
\label{fig:1}       
\end{center}
\end{figure}

Another type of 5D coupling between a hypermultiplet and a vector multiplet
with the same orbifold parities as in the previous example, is
\be
\int d^2\theta \,2H^c(\partial_y-g_{\ga }\epsilon(y)\Sigma) H+
\textup{h.c.},
\la{sup_odd_g}
\ee
where $\epsilon(y)=\partial_y|y|$ is the periodic step-function. Integrating
out the hypermultiplets we get the following contribution to $\Delta$
\be\label{eq:Delta_hyper_charged_2}
         \Delta_{\ga }=\ka_5^2\frac{\gamma}{R^3}\sum_{k=1}^{\infty }
\fr{1}{k^3}e^{-k\pi Rg_{\ga }(\Sigma+\Sigma^+)}\left(1+k\pi Rg_{\ga }
(\Sigma+\Sigma^{+})\right)~,
\ee
where $\ga =N_{H_{\ga }}/2(2\pi )^5$ and $N_{H_{\ga }}$ is the number of
hypermultiplets coupled with the modulus  through the odd coupling
$g_{\ga }\ep (y)$. This expression can e.g. be obtained from formulas given in
\cite{vonGersdorff:2003rq,dudas05} by the 5D lifting of the 1-loop K\"ahler
potential obtained by super-graph method.
In appendix \ref{appA2} we present its derivation in detail.
 One crucial difference to the case discussed previously lies in the fact
that the lowest KK mode is now massless before supersymmetry is broken.
The other relevant difference to eq.\eqref{eq:Delta_hyper_charged_1} is
that it doesn't depend on the axionic modulus $Im(\Sigma)$.

Now we are ready to study the issues of moduli stabilization.
Let us consider the following setup: In addition to a constant (brane)
superpotential $W$ we include the three types of the 1-loop corrections to
the K\"ahler potential discussed in this subsection.
Furthermore we assume the radion  to be the only existing
modulus $\Si $. By a suitable
arrangement of the parameters
$\bar \al =\ka_5^2\al/R^3=2\pi \al/(RM_{\rm Pl})^2$,
$\bar \bt =\ka_5^2\bt /R^3=2\pi \bt/(RM_{\rm Pl})^2$ and
$\bar \ga =\ka_5^2\ga /R^3=2\pi \ga/(RM_{\rm Pl})^2$, as well as
$b=\pi R g_{\bt }/\ka_5=g_{4\bt }\pi RM_{\rm Pl}$ and
$c=\pi R g_{\ga }/\ka_5=g_{4\ga }\pi RM_{\rm Pl}$, one can obtain
a $M_4$ (or dS) minimum for $\Si $.
This is possible since the $\De_{\bt }$ gives a correction opposite to
$\De_{\ga }$ because the imaginary part stabilizes in such a way as to make
some part of the potential negative.
In Fig.\ref{fig:1} we plot such a potential for a specific choice of
parameters. $Re(\Si )$ and $Im(\Si )$ shown in the plot are
dimensionless 4D fields (related to the 5D states according to eq.
(\ref{4Drel5D})).
We tuned $\alpha $ in such a way as to obtain a Minkowski minimum at
$Re(\Si )\simeq 3.5$. Note that the axionic component of
$\Si $ is stabilized as well, in contrast to models previously proposed in
the literature (see e.g. \cite{dudas05}).

Note that there is a class of parameters related to those
of Fig. \ref{fig:1} for which
the stabilization in the $M_4$ vacuum is achieved. To see this, it is
useful to realize that the scalar potential remains invariant (up to the
scale factor) under the following rescalings
\beq
R\to sR~,~~~~(g_{4\bt }~,~ g_{4\ga })\to s^{-1/3}(g_{4\bt }~,~ g_{4\ga })~,~~~~
\Si\to s^{-2/3}\Si ~,
\la{resc}
\eeq
where the scale factor $s$ is a real number.
Note that under this transformation the scalar potential transforms as
$V\to s^2V$ (due to the exponential $e^{\cal K}$ in (\ref{eq:formula})).
However, this change with factor $s^2$ lives the potential at zero
in the vacuum.
With these rescalings the parameters of the model change as
\beq
(\bar \al ~,~\bar \bt ~,~\bar \ga )\to
s^{-2}(\bar \al ~,~\bar \bt ~,~\bar \ga )~,~~~~(b~,~c)\to s^{2/3}(b~,~c)~,
\la{par_resc}
\eeq
allowing to find the class of parameters from those corresponding to
Fig. (\ref{fig:1}) and to have different values of stabilized $\Si $
in the $M_4$ vacuum.

The symmetry displayed in (\ref{resc}) is useful also for discussing
some physical implications. With this rescaling the K\"ahler potential
transforms as
\beq
{\cal K}\to {\cal K}+2\ln s~.
\la{KalTrans}
\eeq
Now it is easy to realize that the particle masses undergo linear
rescalings. Namely, according to (\ref{Mgravitino}) and (\ref{radmasses})
the gravitino and canonically normalized radion component masses change as
\beq
m_{3/2}\to s\cdot m_{3/2}~,~~~~~~m_{\pm }\to s\cdot m_{\pm }~.
\la{masTrans}
\eeq
Therefore, for the class of the parameters related to each other by the
transformation (\ref{par_resc}) the ratio $m_{3/2}/m_{\pm }$ is fixed
\beq
\fr{m_{\pm }}{m_{3/2}}={\cal R}_{\pm }~.
\la{gravRatrad}
\eeq
For the parameters of Fig. \ref{fig:1}
we have ${\cal R}_{+}\simeq 0.17$ and ${\cal R}_{-}\simeq 0.15$.
Thus, with $m_{3/2}\simeq 1$~TeV we have $m_{+}\simeq 170$~GeV and
$m_{-}\simeq 150$~GeV for the two radion scalar component masses.
More generally, since we have a constant superpotential, for the
scalar potential we have $V\propto e^{\cal K}|W|^2$ and comparing with
(\ref{Mgravitino}) we can expect $m_{\pm }\sim m_{3/2}$ unless some
cancellation occurs. Therefore with low energy SUSY
(i.e. $m_{3/2}\sim 1$~TeV) we expect to have radion's two real scalar modes
in the $100~$GeV-$1$~TeV range.
This makes the model  testable in ongoing and future collider
experiments. More details concerning  this issue will be discussed
in sect \ref{sec:spectrum}.

Closing this subsection, let us mention that the conditions in
(\ref{condHighDer1}) for this model can be easily satisfied with low energy
SUSY breaking. This is due the fact that $F$-terms and the radion mass are
close to the TeV scale. Therefore, all effects caused by higher
super-derivative operators can be safely ignored.

\subsection{Non-perturbative stabilization from gaugino condensation}

As we have already mentioned, the flatness of the potential will be lifted
either if the K\"ahler potential receives non cubic (with respect to
$\Si^I+\Si^{I\dagger }$) corrections, or the superpotential has a moduli
dependent
part. In this subsection we will show that the radion stabilization can be
achieved by gaugino condensation.
In this case the effective superpotential will have a moduli dependent
(non perturbative) part. Also the K\"ahler potential gets a non
perturbative correction. The whole effective action still can be
written in a superconformal form
\beq
-3\l e^{-{\cal K}/3}\phi^{\dagger }\phi\r_{D}+\l \phi^3W\r_F +{\rm h.c.}~.
\la{efact}
\eeq
Discussing the gaugino condensation, we follow an effective 4D description
dealing with zero modes of the relevant bulk fields.
Thus, the arguments applied for a pure 4D gaugino condensation
scenario will be appropriate.
For this case the super and K\"ahler potentials include perturbative (p) and
non-perturbative (np) parts \cite{Burgess:1995aa}:
\beq
W=w_p+w_{np}~,~~~e^{-{\cal K}/3}=e^{-{\cal K}_p/3}-ke^{-{\cal K}_{np}/3}~,
\la{efpots}
\eeq
where $k$ is some constant. The action with (\ref{efpots}) is valid below
a scale $\La $ corresponding to energies where the gauge sector becomes
strongly coupled.

{}For demonstrative purposes we will
consider an example with one non Abelian $SU(N)$ YM theory which is
responsible for the gaugino condensation.
Taking the norm function
\beq
{\cal N}(M^I)=M^3-M{\rm Tr}(M_g)^2~, ~~~
{\rm with}~~~M_g=\fr{1}{2}\lam^aM^a~,~~~a=1,\cdots , N^2-1~,
\la{NM}
\eeq
the coupling of the moduli superfield $\Si $ with the gauge field strength
will be
\beq
-\fr{1}{4}\l {\cal N}_{IJ}(\Si ){\cal W}^I{\cal W}^J\r_F+{\rm h.c.}
\to \fr{1}{4}\l \Si {\cal W}^a{\cal W}^a\r_F+{\rm h.c.}
\la{SiWW}
\eeq
The effective 4D superconformal theory with (\ref{efact}), (\ref{SiWW})
possesses Weyl and chiral $U(1)$ symmetries at classical level. However,
one superposition of these two $U(1)$s is anomalous on the
quantum level. Namely, a mixed gauge-chiral anomalous term is
generated and the counter term \cite{Derendinger:1991hq}
\beq
-2c\l \fr{1}{4}\ln \phi {\cal W}^a{\cal W}^a\r_F+{\rm h.c}~
\la{conterm}
\eeq
is needed to take care of the anomaly cancellation. The coefficient $c$
in (\ref{conterm}) is related to the gauge group $b$-factor and is positive
for asymptotically free theories.

Therefore, the coupling of the composite chiral superfield
$U=\lan {\cal W}^a{\cal W}^a\ran $ with moduli and compensating superfields
is given by
\beq
\fr{1}{4}\l (\Si -2c\ln \phi )U \r_F+{\rm h.c.}
\la{consup}
\eeq
Eq. (\ref{consup}) is the starting
point for computing the non perturbative effective action $\Ga (\Si , U)$.
After obtaining the form of $\Ga (\Si , U)$, one can minimize it with
respect to $U$ (determining the condensate $U_0$) and then plug
back the value of $U=U_0$ in $\Ga (\Si , U)$ to derive the effective
action for $\Si $-moduli. In case of a single condensate the non perturbative
K\"ahler and superpotentials are given by \cite{Burgess:1995aa}
\beq
{\cal K}_{np}=\fr{3}{2c}(\Si+\Si^{\dagger })~,~~~~
w_{np}=\tl{w}e^{-\fr{3\Si }{2c}}~.
\la{nonp}
\eeq

The tree level perturbative K\"ahler potential has the form
${\cal K}_p=-3\ln (\Si +\Si^{\dagger })$. Therefore the total
K\"ahler potential
will be
\beq
{\cal K}=-3\ln \l \Si +\Si^{\dagger }-
ke^{-\fr{\Si +\Si^{\dagger }}{2c}}\r ~.
\la{totK}
\eeq
We also assume that the perturbative part of the superpotential is moduli
independent $w_p={\rm const.}$ (often this is the case at tree level).
Note that a non perturbative correction to the K\"ahler potential in the
problem of radion stabilization has not been taken into account so far
in the literature.
There is no physical reason to exclude the $k$-term from considerations.
It also contributes to the kinetic coupling
$(\Si \Si^{\dagger })_D$ for the
moduli field. Since the $k$ is related to a mass scale, we expect that it
will have impact on the stabilized value of $\Si $ moduli.
With $k\neq 0$ the eq. (\ref{totK}) lifts the flatness of the
$ {\cal K}_I{\cal K}^{I\bar J}{\cal K}_{\bar J}-3$ part.
The resulting potential has the form
$$
V=M_{\rm Pl}^4\fr{3|w_p|^2}{(2c)^3}
\fr{\left | \fr{\tl{w}}{w_p}\right |^2(2+r)e^{-2r}+
2\left | \fr{\tl{w}}{w_p}\right |\cos \l \Om -1.5\rho \r e^{-0.5r}-\al }
{(r-\al e^{-r})^2(1+2\al e^{-r}+\al r e^{-r})}e^{-r}~,
$$
\beq
{\rm with}~~~r=\fr{1}{c}{\rm Re}(\Si )~,~~
\rho =\fr{1}{c}{\rm Im}(\Si )~,~~
\al =\fr{k}{2c}~,~~\Om ={\rm Arg}\l \fr{\tl{w}}{w_p}\r ~.
\la{liftfl}
\eeq
%
%
%
\begin{figure}
\rput(7.7,-4.2){({\it i})}
\rput(8,-1.6){({\it iii})}
\rput(6.9,-5.2){({\it ii})}
\rput(13,-2.8){$r$}
\rput(10.2,-0.5){$V/V_0$}
\begin{center}
\leavevmode
\leavevmode
\vspace{2.5cm}
\includegraphics{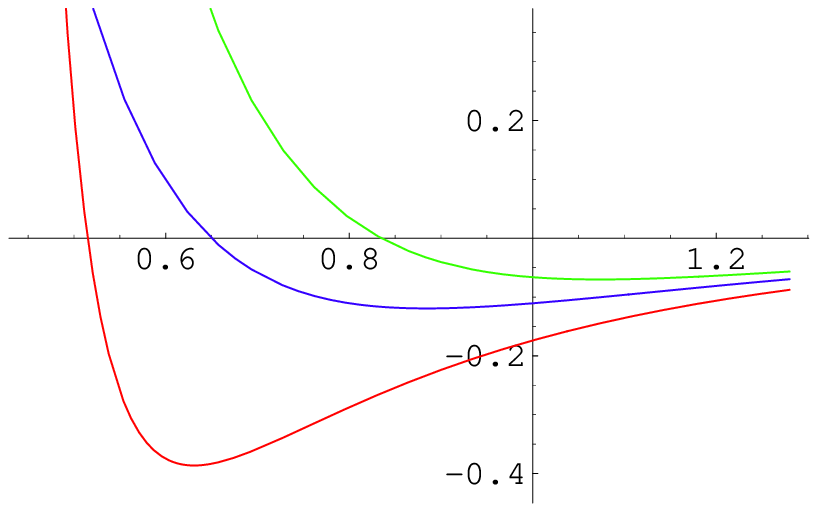}
\end{center}
\vs{3cm}
\caption{
Moduli potential for different cases. $V_0=M_{\rm Pl}^4\fr{3|w_p|^2}{(2c)^3}$,
$\fr{\tl{w}}{w_p}=2$.
({\it i}) $k=0$;
({\it ii})  $\fr{k}{2c}=0.5$;
({\it iii}) $\fr{k}{2c} =-0.5$.
}
\label{fig:2}
\end{figure}
%
%
%
%
{}For $\tl{w}=0$ the radion stabilization is impossible.
Thus, the moduli dependent superpotential plays an important role.
For $\tl{w}\neq 0$ the complex part of the modulus gets fixed as
\beq
\lan \rho \ran =\fr{2}{3}(\Om -\pi )~,
\la{fixcomp}
\eeq
and also the real part is stabilized.
In Fig. \ref{fig:2} we give plots for the cases
$k=0$, $k>0$ and $k<0$, where for the complex part
we have taken the value of eq. (\ref{fixcomp}).
Obviously, the inclusion of the non perturbative
part of the K\"ahler potential introduces significant changes.
In all cases the vacuum energy in the minimum is negative.
In fact, one can easily check that SUSY is unbroken in the minimum:
$D_{\Si }W=0$. The solution of the latter equation, i.e.
\beq
\left | \fr{\tl{w}}{w_p}\right |(1+r)e^{-1.5r}e^{{\rm i}(\Om -1.5\rho )}
+1+\al e^{-r}=0~,
\la{gaugSUSYpres}
\eeq
coincides with the minimum of the potential in (\ref{liftfl}).
Now it is clear why the vacuum ($V=-3e^{\cal K}|W|^2$) is AdS.
An additional contribution to the potential
must be generated \cite{luty99} in order to set it to zero.

We have seen that even though moduli stabilization can be achieved with the
inclusion of non-perturbative effects due to bulk gaugino condensation, this
mechanism still lacks a consistent \emph{uplift} mechanism in order to obtain
Minkowski or de Sitter vacua. Such possibilities will be discussed in the
next section.

\section{Uplift by Boundary Couplings\\
and Stabilization in SUSY Minkowski Vacuum}
\label{sec:treeuplift}

In this section we consider the possibility of uplifting through the boundary
couplings. Also the stabilization in a SUSY preserving Minkowski
vacuum is discussed.

The general form of the scalar ($F$-term) potential is given by

\beq
V=M_{\rm Pl}^4e^{\cal K}\l {\cal K}^{I\bar J}D_IWD_{\bar J}\bar W-3|W|^2\r ~.
\la{totV}
\eeq
Since the first term in (\ref{totV}) is positive definite, we see that
with fixed $W$ the
minimal possible value of $V$ is achieved if the conditions
\beq
D_IW=\pl_IW+{\cal K}_IW=0
\la{SUSYcond}
\eeq
are satisfied. With (\ref{SUSYcond}) we have indeed
$\fr{\pl V}{\pl \Si^I}=-3M_{\rm Pl}^4e^{\cal K}\bar WD_IW=0$ and
therefore the SUSY
preserving solution (\ref{SUSYcond}) is an extremum. However,
the vacuum is AdS ($V=-3M_{\rm Pl}^4e^{\cal K}|W|^2$) unless the
superpotential along this solution is zero (this possibility will be
discussed later on).

It must be noticed that the extremum equation $\fr{\pl V}{\pl \Si^I}=0$
can have solution(s) along which (\ref{SUSYcond}) is not satisfied.
Therefore, there is a chance of having a (meta-stable) vacuum with broken SUSY.
If in this minimum it is possible to set (fine tune) $V=0$, we can have
stabilization in a Minkowski vacuum.

The chiral superfields coming from 5D hypermultiplets can play a
crucial role for uplifting.
Consider the case with one bulk gauge modulus $\Si $ and a set of 4D
chiral multiplets $H_i$ coming from bulk hypers (the $H_i$ are even under
orbifold parity). Assume that the superpotential $W=W(\Si , H_i)$ is arranged
in such a way that  there is a minimum with broken SUSY (i.e. either
$D_{\Si }W$ or at least one $D_{H_i}W$ is non vanishing).
Then along this configuration the potential is
\beq
V=M_{\rm Pl}^4e^{\cal K}\l {\cal K}^{\Si \Si^{\dagger}}|D_{\Si }W|^2
+{\cal K}^{H_iH_i^{\dagger }}|D_{H_i}W|^2-3|W|^2\r ~.
\la{potHdir}
\eeq
If the potential (\ref{potHdir}) fixes
$\Si $ and $H_i$ and if in addition
\beq
 {\cal K}^{\Si \Si^{\dagger}}|D_{\Si }W|^2+
{\cal K}^{H_iH_i^{\dagger }}|D_{H_i}W|^2=3|W|^2
\la{cosmCond}
\eeq
is arranged, then the vacuum is $M_4$.

This can be applied for the case with the gaugino condensation discussed in the previous section. Another interesting example which gives tree level stabilization is with linear superpotential
\beq
W(\Si )=g\Si ~.
\la{linSup}
\eeq
As was discussed in \cite{Correia:2006pj} this superpotential arises
when the compensator hypers are charged under $\Si $ (the $U(1)_R$ is gauged).
Including also the superpotential $W(H)$ for further chiral superfields, the
total superpotential will be
\beq
W=g\Si +W(H)~.
\la{totSup}
\eeq
Restricting ourself to one $H$, coming from the bulk hyper,
the K\"ahler potential is
\beq
{\cal K}=-3\ln (\Si +\Si^{\dagger })-2\ln (1-H^{\dagger }H)~.
\la{SiHKahl}
\eeq
{}For $W$ and ${\cal K}$ given in (\ref{totSup}) and (\ref{SiHKahl}),
one can show that although uplift to $M_4$ can be obtained, not all states
are stabilized. Some of them remain massless. This happens for an arbitrary
form of $W(H)$ (an interesting result). The situation can be improved if
for example in ${\cal K}$
the $\De $ corrections (discussed in sect. \ref{subsec:kaehlerstuff}) are
 included. However,
this goes beyond the tree level analysis and is more involved.
Instead, below we consider a model with a superpotential
combining a gaugino condensation part and the linear one given in
(\ref{linSup}). Although separately both parts lead to the stabilization in
AdS vacua, we will see that together they provide an elegant stabilization
in a SUSY preserving $M_4$ vacuum. Thus, no inclusion of the hypermultiplets
is needed.

The reason we below concentrate on a stabilization in a SUSY preserving Minkowski vacuum is
the following. The phenomenologically viable SUSY breaking can occur in the low energy effective
4D theory by one of the mechanisms widely discussed in the literature. Note that whatever is a
SUSY breaking mechanism, it is preferable to have the gravitino mass much smaller than the mass
of the moduli. In this way the latter will not be destabilized by the SUSY breaking effects. This
also has other advantages especially from cosmological viewpoints \cite{Blanco-Pillado:2005fn}.

\begin{center}
{\bf Stabilization in SUSY Minkowski vacuum}
\end{center}

Now we discuss the issue of the stabilization in a SUSY preserving $M_4$
vacuum. In order to have unbroken SUSY, the conditions (\ref{SUSYcond}) must
be satisfied for all fields. With these conditions, on the other hand, the
potential of
(\ref{totV}) is negative definite unless the superpotential vanishes in the
vacuum.
Thus, in order to have the Minkowski
vacuum we require for the superpotential
\beq
\lan W\ran =0~.
\la{MINcond}
\eeq
This condition in most cases can be achieved by fine tuning. Moreover,
it is very likely that there is another solution with preserving SUSY but
$\lan W\ran \neq 0$. This means that the vacuum (\ref{MINcond}) is a local
minimum.
If the latter is sufficiently long lived we should not worry much about it and
proceed with model building.
Thus, we further invoke the conditions (\ref{SUSYcond}), (\ref{MINcond})
which are equivalent to
\beq
\lan \pl_IW \ran =0~,~~~~\lan W\ran =0~,~~~I=1,2,\cdots
\la{2cond}
\eeq
Of course, one should make sure that the vacuum determined
by these conditions is a minimum. Thus, second derivatives
should be evaluated.

With conditions (\ref{2cond}) we have from (\ref{totV})
\beq
V_{I\bar J}\equiv \fr{\pl^2V}{\pl \Si^I \pl \Si^{\bar J\dagger }}=M_{\rm Pl}^4
e^{\cal K}{\cal K}^{M\bar N}(\pl_M\pl_IW)(\pl_{\bar N}\pl_{\bar J}\bar W)~,~~~
V_{IJ}=V_{\bar I\bar J}=0~.
\la{gen2NDders}
\eeq
The $V_{I\bar J}$ is positive definite and one should make sure that there
is no massless mode(s) in the spectrum.

%
%
%
\begin{figure}
\rput(13,-3.2){$\tl{\Si }$}
\rput(6,-0.7){$V/V_0$}
\begin{center}
\leavevmode
\leavevmode
\vspace{2.5cm}
\includegraphics{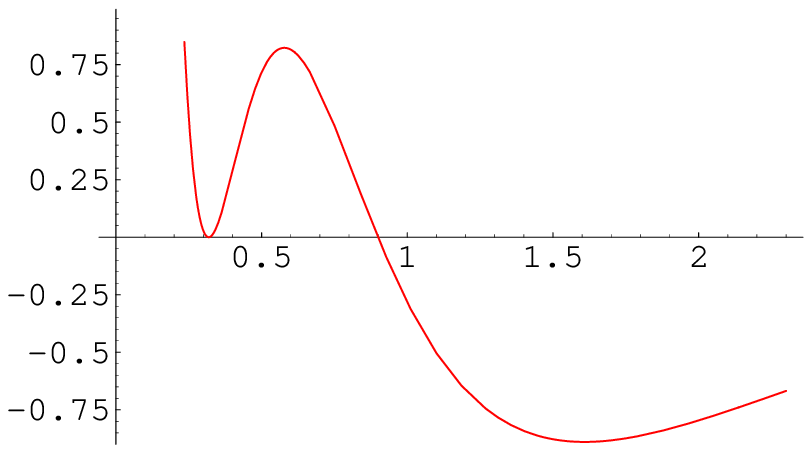}
\end{center}
\vs{3cm}
\caption{
Moduli potential for $\fr{g}{a}=0.02$, $b=2.3$.
$\tl{\Si }\equiv \ln \l2Re(\Si )/3\r $,
$V_0=10^{-5}a^2M_{\rm Pl}^4$.
}
\label{fig:3}
\end{figure}

In case of one modulus, we have only $V_{\Si \Si^{\dagger }}$ and
investigation is simplified.
Consider the superpotential couplings
\beq
W=g\Si +ae^{-b\Si }-\fr{g}{b}\ln \fr{eab}{g}~.
\la{STABmodel}
\eeq
The first term emerges by $U(1)_R$ gauging, while the second term can
come from gaugino condensation as was discussed in the previous section.
All conditions (\ref{2cond}) are satisfied and $\Si $  gets
stabilized at
\beq
\lan \Si \ran =\fr{1}{b}\ln \fr{ab}{g}~.
\la{STABfields}
\eeq
Note that this result does not depend on the form of the K\"ahler potential
because the latter does not appear in the conditions of eq. (\ref{2cond}).
Moreover, the corresponding second derivative
$\pl^2_{\Si }W=gb$ is non zero. Therefore, the desired stabilization in a
SUSY preserving Minkowski vacuum is obtained (imaginary part of
$\Si $ is also fixed). However, there is another
SUSY preserving vacuum along the solution $D_{\Si }W=0$ (with $W\neq 0$)
which is AdS. The potential's profile for
one possible choice of parameters (which gives $Im(\Si )=0$) is shown in
Fig. \ref{fig:3}.
As was analyzed in early times \cite{Weinberg:1982id}  the Minkowski vacuum
can be  fairly long lived, i.e. meta-stable
(for a recent discussion see \cite{Intriligator:2006dd}).

Before concluding this section, it is useful to present some expressions
for the case of two fields.
Consider one gauge modulus $\Si $
and a 4D chiral superfield $H$ (of positive parity) coming from a bulk
hypermultiplet.
In general, the superpotential $W$ is a function of both superfields
$W=W(\Si , H)$.
For bulk modulus and hyper,
the tree level K\"ahler potential is given by (\ref{genK}) and therefore
${\cal K}_{\Si H^{\dagger }}=0$. From (\ref{gen2NDders}) we have
$$
V_{\Si \Si^{\dagger }}\equiv \fr{\pl^2V}{\pl \Si \pl \Si^{\dagger }}=
M_{\rm Pl}^4e^{\cal K}\l {\cal K}^{\Si \Si^{\dagger }}|\pl^2_{\Si}W|^2 +
 {\cal K}^{HH^{\dagger }}|\pl_{\Si}\pl_HW|^2\r ~,
$$
$$
V_{\Si H^{\dagger }}\equiv \fr{\pl^2V}{\pl \Si \pl H^{\dagger }}=
\l \fr{\pl^2V}{\pl \Si^{\dagger } \pl H}\r^*=
M_{\rm Pl}^4e^{\cal K}\l {\cal K}^{\Si \Si^{\dagger }}
(\pl^2_{\Si}W)(\pl_{\Si^{\dagger }}\pl_{H^{\dagger }}\bar W) +
 {\cal K}^{HH^{\dagger }}(\pl_{\Si } \pl_HW)(\pl^2_{H^{\dagger }}\bar W)\r ~,
$$
\beq
V_{HH^{\dagger }}\equiv \fr{\pl^2V}{\pl H \pl H^{\dagger }}=
M_{\rm Pl}^4e^{\cal K}\l {\cal K}^{\Si \Si^{\dagger }}|\pl_{\Si }\pl_HW|^2 +
 {\cal K}^{HH^{\dagger }}|\pl^2_HW|^2\r ~,
\la{2NDders}
\eeq
and all remaining second derivatives vanish.
In order the vacuum to be a local minimum all eigenvalues of the matrix

\begin{equation}
\begin{array}{cccc}
 & {\begin{array}{cccc}
      \hs{-0.7cm} &\hs{0.5cm}
~ &\hs{0.5cm}~
\end{array}}\\ \vspace{1mm}

\begin{array}{c}
 \\  \\
 \end{array}\!\!\!\!\!\hs{-0.2cm} &{\left(\begin{array}{cccc}

 4V_{\Si \Si^{\dagger }} & ~0  &V_{\Si H^{\dagger }}+V_{\Si^{\dagger }H}~&
{\rm i}(V_{\Si^{\dagger }H}-V_{\Si H^{\dagger }})
\\
 0 &~4V_{\Si \Si^{\dagger }} &{\rm i}(V_{\Si H^{\dagger }}-V_{\Si^{\dagger }H})
&V_{\Si H^{\dagger }}+V_{\Si^{\dagger }H}
 \\
 V_{\Si H^{\dagger }}+V_{\Si^{\dagger }H} &
{\rm i}(V_{\Si H^{\dagger }}-V_{\Si^{\dagger }H}) &
~ 4V_{HH^{\dagger }} &0
\\
{\rm i}(V_{\Si^{\dagger }H}-V_{\Si H^{\dagger }})&
V_{\Si H^{\dagger }}+V_{\Si^{\dagger }H}
 &0 & 4V_{HH^{\dagger }}

\end{array}\right)}~,
\end{array}  \!\!  ~~~~~
\label{2DERmatr}
\end{equation}
must be positive. This condition should be required together with
(\ref{2cond}). A more detailed investigation will be required if by chance
some of the eigenvalues vanish.

In the concrete case with a superpotential  given by $W=W(\Si )+W(H)$ all
off diagonal elements in (\ref{2DERmatr}) vanish and the eigenvalues are
positive semidefinite. Thus, the model building is extremely simplified.

\section{Mass Spectrum and Phenomenological Implications}\label{sec:spectrum}

Now we present mass formulas for the stabilized modulus and also for the 4D
gravitino. We consider the case of one modulus without hypers.
Inclusion of the hypers and boundary chiral multiplets and a generalization
of the corresponding expressions are straightforward.

The canonically normalized scalar component
$\Si_c=\fr{1}{\sq{2}}(\Si_c^R+{\rm i}\Si_c^I)$ of the modulus superfield
is related to $\Si $ by the differential relation
\beq
d\Si_c=M_{\rm Pl}{\cal K}_{\Si \Si^{\dagger }}^{1/2}d\Si ~.
\la{canSi}
\eeq
After $Re(\Si )$ and $Im(\Si )$ get stabilized, in general $\Si_c^R-\Si_c^I$
mixing occurs and the squared mass matrix is given by
$$
\begin{array}{ccc}
 & {\begin{array}{ccc}
      \hs{-0.7cm} &\hs{0.5cm}
~ &\hs{0.5cm}~
\end{array}}\\ \vspace{1mm}
\begin{array}{c}
 \\  \\
 \end{array}\!\!\!\!\!\hs{-0.2cm} &{\left(\begin{array}{cc}

 V_{RR} & ~V_{RI}
\\
 V_{RI} &~V_{II}

\end{array}\right)}~,~~~~{\rm with}
\end{array}  \!\!  ~~~~~
$$
\beq
V_{RR}+V_{II}=\fr{2}{M_{\rm Pl}^2{\cal K}_{\Si \Si^{\dagger}}}
V_{\Si \Si^{\dagger}}~,~
~~V_{RR}V_{II}-V_{RI}^2=\fr{1}{M_{\rm Pl}^4{\cal K}_{\Si \Si^{\dagger}}^2}
\l V_{\Si \Si^{\dagger}}^2-V_{\Si \Si}V_{\Si^{\dagger} \Si^{\dagger}}\r ~.
\label{MasmatrSic}
\end{equation}
Here and below we omit $\lan \cdots \ran $ symbols keeping in mind that
all quantities are evaluated in the vacuum.
The eigenvalues of the mass matrix in (\ref{MasmatrSic}) are given by
\beq
m_{\pm }^2=\fr{1}{M_{\rm Pl}^2{\cal K}_{\Si \Si^{\dagger}}}
\l V_{\Si \Si^{\dagger}} \pm |V_{\Si \Si}| \r ~.
\la{radmasses}
\eeq
These mass scales are phenomenologically constrained. Namely, the radion's
scalar component's masses should not be less than few$\cdot 10^{-3}$~eV
\cite{Adelberger:2003zx} in order to avoid an unacceptable large deviation
from the Newton's 4D gravitational potential. The source for the deviation
is the exchange of light scalar modes.

In case of an unbroken SUSY in $M_4$ vacuum (i.e. $\pl_{\Si }W=W=0$)
we have from (\ref{gen2NDders})
$V_{\Si \Si }=0$ and both components of the $\Si_c$ have the same mass:
\beq
m^2_{\Si_c}=\fr{1}{M_{\rm Pl}^2{\cal K}_{\Si \Si^{\dagger}}}
V_{\Si \Si^{\dagger}}
=M_{\rm Pl}^2e^{\cal K}|{\cal K}^{\Si \Si^{\dagger }}\pl^2_{\Si }W|^2~.
\la{SUSYradmass}
\eeq
Of course, for this case the fermionic partner of $\Si^c$, the modulino,
has the same mass and the gravitino mass $m_{3/2}$ vanishes.
{}For the example presented in sect. \ref{sec:treeuplift} with
superpotential given in eq. (\ref{STABmodel}) we have
\beq
m_{\Si_c}=\fr{g}{3}\l 2b\ln \fr{ab}{g}\r^{1/2}M_{\rm Pl}~.
\la{STABM4mass}
\eeq
Clearly, by an appropriate selection of $g$ (not fixed yet) it is possible
to get a phenomenologically acceptable mass value.

In case of broken SUSY the gravitino obtains mass. With the $F$-term
SUSY breaking (which we consider in this paper) in $M_4$ vacuum we have
\beq
m_{3/2}=M_{\rm Pl}|W|e^{{\cal K}/2}~.
\la{Mgravitino}
\eeq
The masses of the radion's scalar components should be evaluated by
eq. (\ref{radmasses}). The fermionic component of the superfield $\Si $
is a goldstino absorbed into the  $1/2$-spin mode of the gravitino. (The
$F_{\phi }$ component of a compensator is zero and the only source here for
the SUSY breaking is a non zero $F^{\Si }$ term).

{}For the example presented in sect. \ref{subsec:kaehlerstuff}, the value of
the constant superpotential $W$ is not fixed. Thus, this scenario allows
to select $W$ in such a way as to keep $m_{3/2}$ in the TeV range. This is
one advantage of this model over scenarios where  uplifting and SUSY
breaking occur by a (fixed) $D$-term.
For the choice of parameters corresponding to Fig. \ref{fig:1} for the
gravitino and radion's scalar component masses we have respectively
\beq
m_{3/2}\simeq 5\cdot 10^{-2}M_{\rm Pl}|W|~,~~~
m_{+}\simeq 10^{-2}M_{\rm Pl}|W|~,~~~
m_{-}\simeq 8\cdot 10^{-3}M_{\rm Pl}|W|~.
\la{grtinorads}
\eeq
With $|W|\sim 10^{-15}$ all these states lie near the TeV scale.
As was already pointed out in sect. \ref{subsec:kaehlerstuff}, for this
model we naturally expect $m_{\pm }\sim m_{3/2}$, unless
due to a cancellation $m_{-}$ is much smaller then $m_{3/2}$.
This can happen with $V_{\Si \Si^{\dagger }}\simeq V_{\Si \Si}$ (see
eq. (\ref{radmasses})) realized when the potential weakly depends
on the axionic component of the radion, i.e. with strongly suppressed
$\bar \bt $.
However, in this case apart from problem of axionic component's stabilization
it is difficult (if not impossible) to stabilize also the real component
in the $M_4$ vacuum.
Thus, we conclude that with  low energy SUSY breaking radion scalar components
masses are expected to be in the $100$~GeV-$1$~TeV energy range.
This opens up a window for a collider physics with an interesting
phenomenological signature \cite{Han:1998sg}.

\section{Summary}\label{sec:conc}

In this paper we have presented several new examples of radion stabilization
and found possibilities of uplifting to Minkowski vacuum which is a highly
non trivial task.
We have applied an effective 4D superfield description which makes
investigations rather simple and provides an excellent playground for studying
various ideas and for phenomenological model building. The formalism
also allows  us to test various mechanisms of moduli stabilization.
This  can for example be useful for a more detailed study
of inflation together with radion stabilization,
but also for a more general moduli dynamics in the early Universe. These
issues will be discussed elsewhere.

\vs{0.5cm}

\hs{-0.7cm}{\bf Acknowledgments}

\vs{0.2cm}
\hs{-0.7cm}We thank Emilian Dudas, Boris Kors, Sergei Kuzenko
and Marek Olechowski for useful discussions.
The research of F.P.C. is supported by Funda\c c\~ ao para a
Ci\^ encia e a Tecnologia (grant SFRH/BPD/20667/2004).

\appendix

\renewcommand{\theequation}{A.\arabic{equation}}\setcounter{equation}{0}
\section{One-loop corrections to the K\"ahler potential
from bulk hypermultiplets}\label{appA}

\subsection{Case with even coupling}\label{appA1}

First we derive a 1-loop correction to the K\"ahler potential due to
hypermultiplets coupled with the vector modulus through even coupling
[eq. (\ref{eq:hyper_phys_charged_1})].
There are several ways of deriving eq.\eqref{eq:Delta_hyper_charged_1}.

\begin{center}
{\bf Calculation from prepotential}
\end{center}

First we will take  the following approach: we calculate first the prepotential $\mcF(\Sigma)$ of $\mcN=2$ supergravity compactified on the $S^1$ and then
use the well-known expression (see e.g. \cite{Cremmer:1984hj})
\be
\mcK=-\ln\left(-2(\mcF+\mcF^+)+
(\mcF_I+\mcF_{\bar I}^+)(\Sigma^I+\Sigma^{{\bar I}+})\right),
\ee
to obtain the K\"ahler potential $\mcK(\Sigma,\Sigma^+)$ from $\mcF(\Sigma)$.
The reason one can use the $S^1$ result also in the orbifold case is due to
the fact that the relevant KK spectrum is essentially the same in both cases.

Recall that the prepotential $\mcF(\Sigma)$ determines the holomorphic gauge
couplings through its second derivative:
\be
         -\frac{1}{4}\int d^2\theta\mcF''(\Sigma)\mcW\mcW+\textup{h.c.}
\ee
Knowing the $\Sigma$ dependent 1-loop correction to the gauge coupling we
will be able to obtain $\mcF$ and therefore to determine the K\"ahler
potential. It is not difficult to find out that
\be
                   \mcF''(\Sigma)=B\ln2\sinh(\pi R g_{\bt }\Sigma)=
B\left[\pi Rg_{\bt }\Sigma-\sum_{k=1}^{\infty}
\frac{e^{-k2\pi Rg_{\bt }\Sigma}}{k}\right],
\ee
with $B=2\pi^2 g^2_\beta /R$. A double integration gives
\be
                   \mcF=B\left[\frac{\pi Rg_{\bt }}{6}\Sigma^3-
\sum_{k=1}^{\infty}\frac{e^{-k2\pi Rg_{\bt }\Sigma}}{(2\pi Rg)^2k^3}\right].
\la{F}
\ee
The first term (on the r.h.s.) amounts to a renormalization of the
Chern-Simons term $\sim M^3$ in the norm-function. The second term in
$\mcF$ is the
correction ($\De \mcF$) we are searching for. It gives
\be
              \Delta_{\bt }  =-2(\De \mcF+\De \mcF^+)+
(\De \mcF'+{\De \mcF'}^+)(\Sigma+\Sigma^{+})~.
\la{DeF}	
\ee
Using in (\ref{DeF}) the second term of (\ref{F}) we arrive at the expression
given in (\ref{eq:Delta_hyper_charged_1}).
Next we present a derivation by a different method.

\begin{center}
{\bf Calculation in terms of $N=1$ superfields}
\end{center}

Since we are able to write our 5D action in terms of 4D $N=1$ superfields,
we can take an advantage and use some results existing in 4D constructions.
If the Lagrangian couplings are written in terms of 4D superfields as
\beq
\int d^4\te K(\hat{\ov{\phi }}, \hat{\phi })+\int d^2\te W_0(\phi)
+{\rm h.c}+\cdots
\la{4Dcoupl}
\eeq
the
1-loop correction to $K$ (this should not be confused with ${\cal K}$ used
for the K\"ahler potential in the effective 4D superconformal description) is
given by\footnote{One can also proceed with technics presented in
\cite{Buchbinder:1998qv} evaluating the $\log $ of a super determinant
which is written as (\ref{DeKSi}).} \cite{Brignole:2000kg}
\beq
\De K=\fr{1}{(2\pi )^4}\int \fr{d^4p}{p^2}\left [
{\rm Tr}\log \hat{K}+\fr{1}{2}{\rm Tr}\log \l p^2+
\hat{K}^{-1}\hat{\ov{W}_0}\hat{K}^{-1T}\hat{W}_0\r  -
{\rm Tr}\log\l \hat{H}p^2+\hat{S}\r
\right ]~,
\la{DelK}
\eeq
where
\beq
\hat{K}_{\ov{I}J}=K_{\ov{I}J}(\hat{\ov{\phi }}, \hat{\phi })~,~~~
(\hat{W}_0)_{IJ}=(W_0)_{IJ}(\hat{\phi })~,~~~{\rm etc}.
\la{defs}
\eeq

The 5D Lagrangian of hypermultiplets coupled to the modulus by an even
coupling is given by \cite{us04a}
\beq
{\cal L}(H)=2\int d^4\te {\mathbb W}_y(H^{\dagger }H+H^{c\dagger }H^c)+
\int d^2\te \l 2H^c\pl_yH-g\Si (H^2-H^{c2})\r +{\rm h.c.}
\la{evCoup}
\eeq
where we have set $V=0$, and all objects are five dimensional.
(Here and in the equations below instead of $g_{\bt }$ we use $g$
to make expressions more compact).
In order to write this Lagrangian in the form of (\ref{4Dcoupl})
here we perform the KK decomposition and integrate over the fifth dimension
$y$. Thus we will be able to apply the expression (\ref{DelK}) for each
KK state, summing at the end over the full tower.
For chiral superfields with positive and negative orbifold parities
respectively we will
have
\beq
H=\fr{1}{\sq{4\pi R}}H^{(0)}+
\fr{1}{\sq{2\pi R}}\sum_{n=1}^{\infty }H^{(n)}\cos \fr{ny}{R}~,~~~
H^c=\fr{1}{\sq{2\pi R}}\sum_{n=1}^{\infty }H^{c(n)}\sin \fr{ny}{R}~
\la{KKdec}
\eeq

$$
\int dy{\cal L}(H)=\int d^4\te {\mathbb W}_y\l
\sum_{n=0}^{\infty } H^{(n)\dagger }H^{(n)}+
\sum_{n=1}^{\infty } H^{c(n)\dagger }H^{c(n)}\r
$$
\beq
-\int d^2\te \l \fr{1}{R}\sum_{n=0}^{\infty }nH^{c(n)}H^{(n)}+
\fr{g}{2}\Si \sum_{n=0}^{\infty }(H^{(n)})^2-
\fr{g}{2}\Si \sum_{n=1}^{\infty }(H^{c(n)})^2\r +{\rm h.c.}
\la{KKL}
\eeq
Comparing this with eqs (\ref{4Dcoupl}), (\ref{defs}) we see that $\hat{K}$
and $\hat{W}_0$ are matrices in KK space:

\begin{equation}
\begin{array}{ccc}
 & {\begin{array}{ccc}
\hs{0cm}H^{(0)}\hspace{0.1mm} & H^{(n)}
 &\hs{0.1mm} H^{c(n)}
\end{array}}\\ \vspace{1mm}
\hat{K}=
\begin{array}{c}
H^{(0)\dagger }\vs{0.1cm} \\ H^{(n)\dagger } \vs{0.1cm} \\ H^{c(n)\dagger }
 \end{array}\!\!\!\!\!\hs{-0.2cm} &{\left(\begin{array}{ccc}

 {\mathbb W}_y &0  & 0
\\
 0 &{\mathbb W}_y &0
 \\
0& 0 &{\mathbb W}_y

\end{array}\right)}~,~~~
\end{array}
\begin{array}{ccc}
 & {\begin{array}{ccc}
\hs{0.2cm}H^{(0)} &\hs{0.1cm} H^{(n)} & \hs{0.1cm} H^{c(n)}
\end{array}}\\ \vspace{1mm}
\hat{W}=
\begin{array}{c}
H^{(0)} \\ H^{(n)} \\ H^{c(n)}
 \end{array}\!\!\!\!\!\hs{-0.2cm} &{\left(\begin{array}{ccc}

 -g\Si &~0 &~ 0
\\
 0 &~ -g\Si &~\fr{n}{R}
 \\
 0 &~\fr{n}{R}&~ g\Si

\end{array}\right)}~.
\end{array}  \!\!
\label{KWmatr}
\end{equation}
and
\begin{equation}
\begin{array}{ccc}
 & {\begin{array}{ccc}
      \hs{-0.7cm} &\hs{0.5cm}
~ &\hs{0.5cm}~
\end{array}}\\ \vspace{1mm}
{\cal M}^2_H\equiv \hat{K}^{-1}\hat{\ov{W}}_0\hat{K}^{-1T}\hat{W}_0=
\begin{array}{c}
 \\  \\
 \end{array}\!\!\!\!\!\hs{-0.2cm} &{\left(\begin{array}{ccc}

 |\om |^2 & ~0  &~0
\\
 0 &~n^2+|\om |^2 &~-n(\om^{\dagger }-\om )
 \\
 0 &~ -n(\om -\om^{\dagger } ) &~ n^2+|\om |^2

\end{array}\right)\fr{1}{R^2{\mathbb W}_y^2}}~,
\end{array}  \!\!  ~~~~~
\label{matr}
\end{equation}
where $\om =gR\Si $.
Eigenvalues of ${\cal M}^2_H$ are
\beq
\fr{|\om |^2}{R^2{\mathbb W}_y^2}~,~~~
\fr{1}{R^2{\mathbb W}_y^2}\l (n\pm \fr{1}{2}|\om^{\dagger }-\om |)^2+
\fr{1}{4}(\om +\om^{\dagger })^2\r ~,~~~n=1,2,\cdots
\la{spectrum}
\eeq
Therefore,
\beq
\De K(\Si )=\fr{1}{32\pi^2}\int dp^2{\rm Tr}\log (p^2+{\cal M}_H^2)=
-\fr{1}{32\pi^2}\fr{1}{R^2{\mathbb W}_y^2}\int \fr{dt}{t^2}{\rm tr}
e^{-R^2{\mathbb W}_y^2{\cal M}_H^2t}\equiv \fr{1}{32\pi^2}
\fr{1}{R^2{\mathbb W}_y^2}{\cal I}~.
\la{DeKSi}
\eeq
where we have dropped  $\Si $-independent parts. We
furthermore  evaluate ${\cal I}$. Using Poisson resummation we have
\beq
{\cal I}=\hs{-0.2cm}\int \fr{dt}{t^2}e^{-(\om +\om^{\dagger })^2t/4}
\hs{-0.2cm}\sum_{n=-\infty }^{+\infty }e^{(n+|\om^{\dagger }-\om |^2/2)^2t}=
\sq{\pi }\hs{-0.2cm}\int \fr{dt}{t^{5/2}}e^{-(\om +\om^{\dagger })^2t/4}
\hs{-0.2cm}\sum_{k=-\infty }^{+\infty }
e^{-\pi^2k^2/t}e^{{\rm i}\pi k|\om^{\dagger }-\om |}~.
\la{calcI}
\eeq
The $k=0$ term gives
\beq
{\cal I}_{k=0}=\sq{\pi }\int \fr{dt}{t^{5/2}}
e^{-(\om +\om^{\dagger })^2t/4}\sim
|\om +\om^{\dagger }|^3\La^3\sim (\Si +\Si^{\dagger })^3\La^3~,
\la{Ik0}
\eeq
where $\La $ is the UV cut off ($t_{UV}\sim 1/\La^2$). This contribution
renormalizes the tree level K\"ahler potential.
For us contributions coming from $k\neq 0$ terms are interesting.
These contributions are finite for $\La \to \infty $.
Using the integration  formula
\beq
\int_0^{\infty } \fr{dt}{t^{5/2}}e^{-at-b/t}=
\fr{\sq{\pi }}{2b^{3/2}}(1+2\sq{ab})
e^{-2\sq{ab}}~,
\la{52int}
\eeq
and the abbreviation
\beq
2gR\hat{\Si }=
|\om +\om^{\dagger }|+{\rm i}(\om^{\dagger }-\om )~,
\la{abrSi}
\eeq
from (\ref{DeKSi}) and (\ref{calcI}) we get finally
\beq
\De K(\Si )=-\fr{1}{64\pi^4}\fr{1}{R^2{\mathbb W}_y^2}
\sum_{k=1}^{\infty }\fr{1}{k^3}
\l e^{-2\pi kgR\hat{\Si }}+e^{-2\pi kgR\hat{\Si }^{\dagger }}\r
\l 1+\pi kgR(\hat{\Si }+\hat{\Si }^{\dagger })\r ~.
\la{finDeK}
\eeq
Recalling that we are dealing only with the zero mode of $\Si $ we can write
$\De K(\Si )=\fr{1}{2\pi R}\int dy \De K(\Si )$
and we  precisely get the expression of eq. (\ref{eq:Delta_hyper_charged_1})
for $\De_{\bt }$ (with taking into account that the 4D $\De $ is related
to 5D one as $\De =\ka_5^2\De_5$).

\subsection{Case with odd coupling}\label{appA2}

In this case, the relevant hypermultiplet 5D Lagrangian couplings are
\beq
{\cal L}(H)=2\int d^4\te {\mathbb W}_y(H^{\dagger }H+H^{c\dagger }H^c)+
2\int d^2\te H^c\l \pl_y-g\Si \ep(y)\r H +{\rm h.c.}
\la{oddCoup}
\eeq
To obtain eq.\eqref{eq:Delta_hyper_charged_2} we will have to resort to a
different, less elegant, technique.
Note that the eigenvalues of the matrix
${\cal M}^2_H\equiv \hat{K}^{-1}\hat{\ov{W}}_0\hat{K}^{-1T}\hat{W}_0$
coincide with the KK mass spectrum.
Therefore, instead of KK decomposition we will find the mass eigenvalues
by solving the 5D eigenstate equations with appropriate boundary conditions
for hypers' component chiral superfields.

In the $D$-term part of (\ref{oddCoup}), we use the replacement
$d^2\bar \te^2=-\fr{1}{4}\bar D^2$ and treat the ${\mathbb W}_y$ and $\Si $ as
backgrounds\footnote{We assume that higher super-derivative terms have
negligible effect on a result. This assumption must be justified and
we will discuss this issue at the end of this appendix.}:
$D^2{\mathbb W}_y=\bar D^2{\mathbb W}_y=D^2\Si =\bar D^2\Si =0$ etc.
Then from (\ref{oddCoup}) we can write equations of motion for superfields
$H, H^c$
$$
-\fr{1}{4}{\mathbb W}_y\bar D^2H^{\dagger }-(\pl_y+g\Si \ep(y))H^c=0~,
$$
\beq
-\fr{1}{4}{\mathbb W}_y\bar D^2H^{c\dagger }+(\pl_y-g\Si \ep(y))H=0~.
\la{eqsH}
\eeq
With change of variables
\beq
H=e^{g\Si |y|}\Phi ~,~~~~H^c=e^{-g\Si |y|}\Phi^c~,
\la{Hphi}
\eeq
using the identity
$D^2\bar D^2=\bar D^2D^2+8{\rm i}\bar D\si^{\mu }D\pl_{\mu }+16\square_4$
and making some manipulations, from (\ref{eqsH}) we obtain
$$
{\mathbb W}_y^2m^2\Phi +e^{-g(\Si +\Si^{\dagger })|y|}\pl_y\l
e^{g(\Si +\Si^{\dagger })|y|}\pl_y\Phi \r =0~,
$$
\beq
{\mathbb W}_y^2m_c^2\Phi^c +e^{g(\Si +\Si^{\dagger })|y|}\pl_y\l
e^{-g(\Si +\Si^{\dagger })|y|}\pl_y\Phi^c \r =0~,
\la{eqsPhi}
\eeq
where we have used $\square_4\Phi =m^2\Phi $ and
$\square_4\Phi^c =m_c^2\Phi^c $. The (\ref{eqsPhi}) should be solved
by boundary conditions
\beq
\Phi (-y)=\Phi (y)~,~~\Phi^c (-y)=-\Phi^c (y)~,~~
H(y+2\pi R)=H(y)~,~~H^c(y+2\pi R)=H^c(y)~,
\la{bcHPhi}
\eeq
which also give discrete values of $m^2$ and $m_c^2$.
In the interval $0<y<\pi R$ we have the solution
\beq
\Phi =e^{-g(\Si +\Si^{\dagger })y/2}\l Ae^{\om y}+Be^{-\om y}\r ~,~~~
{\rm with}~~~\om =\l \fr{g^2}{4}(\Si+\Si^{\dagger })^2-
{\mathbb W}_y^2m^2\r^{1/2}~,
\la{solPhi}
\eeq
and for $-\pi R<y<0$
\beq
\Phi =e^{g(\Si +\Si^{\dagger })y/2}\l Ae^{-\om y}+Be^{\om y}\r ~.
\la{solPhineg}
\eeq
Using (\ref{eqsPhi}) and the boundary conditions we get
\beq
\om R={\rm i}n~.
\la{discr}
\eeq
Thus, for the mass eigenvalues we get
\beq
m_n^2=\fr{1}{R^2{\mathbb W}_y^2}\l n^2+\fr{1}{4}g^2R^2(\Si +\Si^{\dagger })^2\r ~.
\la{mn}
\eeq
Similarly, we obtain
\beq
(m_c^2)_n=\fr{1}{R^2{\mathbb W}_y^2}\l n^2+\fr{1}{4}g^2R^2
(\Si +\Si^{\dagger })^2\r ~.
\la{mcn}
\eeq
Therefore, we can write
\beq
\De K(\Si )=-\fr{1}{32\pi^2}\fr{1}{R^2{\mathbb W}_y^2}\int \fr{dt}{t^2}
\sum_{n=-\infty }^{\infty }e^{-R^2{\mathbb W}_y^2m_n^2t}=
-\fr{1}{32\pi^2}\fr{1}{R^2{\mathbb W}_y^2}\int \fr{dt}{t^2}
\sum_{n=-\infty }^{\infty }e^{-[n^2+g^2R^2(\Si +\Si^{\dagger })/4]t}~.
\la{oddK}
\eeq
Making similar steps as in the previous subsection, we finally get
\beq
\De K(\Si )=-\fr{1}{32\pi^4}\fr{1}{R^2{\mathbb W}_y^2}
\sum_{k=1}^{\infty }\fr{1}{k^3}
\l 1+\pi kgR|\Si +\Si^{\dagger }|\r e^{-\pi kgR|\Si +\Si^{\dagger }|}~,
\la{finKodd}
\eeq
which for $\De_{\ga }$ gives the expression given in
(\ref{eq:Delta_hyper_charged_2}).

In our analysis we have ignored the higher superderivative terms which are
powers of $D^2\Si $, $D^2{\mathbb W}_y$,  $\bar D^2D^2{\mathbb W}_y$ etc.
In general, the solution  eqs. (\ref{solPhi}), (\ref{solPhineg}) receive
the following type of corrections
\beq
\sum a_{ij}(D^2\Si)^i(D^2{\mathbb W}_y)^j+
\sum b_i(\bar D^2D^2{\mathbb W}_y)^i+\cdots
\la{corHighDer}
\eeq
because of modification of (\ref{eqsH}), (\ref{eqsPhi}).
One can check out that these corrections can be safely ignored if the
following conditions are satisfied
\beq
m^2, m_c^2\gg \fr{({\mathbb W}_y)_D}{{\mathbb W}_y}~,~~~
m^2, m_c^2\gg g_52\pi R|\Si_5|m_{\Si }^2=g_42\pi RM_{\rm Pl}|\Si |m_{\Si }^2~.
\la{condHighDer}
\eeq
{}For ${\mathbb W}_y$ we have \cite{Correia:2006pj}
\beq
{\mathbb W}_y=\l \fr{\ka_5^{-2}(\tl{\cal N}+\De )}{h^{\dagger }h}\r^{1/3}=
M_{\rm Pl}\fr{e^{-{\cal K}/3}}{(\phi^{\dagger }\phi )^{1/2}}~,
\la{exprWy}
\eeq
and therefore
\beq
\fr{({\mathbb W}_y)_D}{{\mathbb W}_y}=\left | \fr{1}{3}{\cal K}_IF^I+
\fr{1}{2}\fr{F_{\phi }}{\phi }\right |^2~.
\la{WyD}
\eeq
{}For the conditions (\ref{condHighDer}) to be satisfied it is enough
to take the zero mode masses $m_0^2$ and $(m^2_c)_0$. Thus, taking
into account (\ref{mn}), (\ref{mcn}), and (\ref{WyD}), eq.
(\ref{condHighDer}) can be rewritten as
\beq
\fr{g_4^2}{4{\mathbb W}_y^2}(\Si +\Si^{\dagger })^2M_{\rm Pl}^2\gg
\left | \fr{1}{3}{\cal K}_IF^I+\fr{1}{2}\fr{F_{\phi }}{\phi }\right |^2~,~
g_42\pi RM_{\rm Pl}|\Si |m_{\Si }^2~.
\la{condHighDer1}
\eeq
Now it is easy to see that these conditions restrict the amount of SUSY
breaking and the value of the radion mass. For a concrete model one can easily
check out whether these conditions are met or not.

\end{document}